\documentclass[pra,aps,twocolumn,showpacs,amsmath,amssymb,nofootinbib]{revtex4}
\usepackage{epsfig}
\usepackage{graphics}
\usepackage{bm}
\usepackage{amssymb,amsmath}
\usepackage{graphicx}

\newcommand{\ket}[1]{|#1\rangle}             
\newcommand{\bra}[1]{\langle #1|}            

\newcommand{\be}{\begin{equation}}
\newcommand{\ee}{\end{equation}}
\newcommand{\beqa}{\begin{eqnarray}}
\newcommand{\eeqa}{\end{eqnarray}}
\newcommand{\beq}{\begin{equation}}
\newcommand{\eeq}{\end{equation}}
\def\sa{\textsf}

\begin{document}
\title{Low velocity limits of cold atom clocks}
\author{J. Mu\~noz}
\email{josemunoz@saitec.es}
\author{I. Lizuain}
\email{dirk_seidel@ehu.es}
\author{J. G. Muga}
\email{jg.muga@ehu.es}
\affiliation{Departamento de Qu\'{\i}mica-F\'{\i}sica, UPV-EHU, Apartado Postal 644, 48080 Bilbao, Spain}

\begin{abstract}
Fundamental low-energy limits to the accuracy of quantum clock and stopwatch
models in which the clock hand motion is activated by the presence of a
particle in a region of space have
been studied in the past, but their relevance for actual atomic clocks
had not been assessed. In this work we address the 
effect of slow atomic quantum motion on Rabi and Ramsey resonance
fringe patterns, 
as a perturbation of the results based on classical atomic motion.
We find the dependence of the fractional error of the corresponding atomic clocks on the atomic velocity and
interaction parameters.        
\end{abstract}
\maketitle
%
\section{Introduction}
There is a sizable amount of work devoted to ``quantum clocks''
in which a clock ``hand'' variable runs 
during the passage of a particle through a region of space
\cite{Baz,Ry,Peres80,Bu83,L93,L94,Peres95,Bra97,L98,AOPRU,CR00,Alonso,S07}, see an extensive review in  \cite{SAE}. Much of this work is related to the ``Larmor clock'' 
for an electron crossing a potential barrier and the determination of tunneling times \cite{Bu08}, but even the timing of a simple freely-moving particle poses  intriguing questions \cite{dwell}.   
In principle, and from a classical-mechanical perspective,
reading the hand after the particle has crossed the region 
(the particle may also be put at time $t=0$ 
{\em within} the region) should 
provide a measurement of the time that the particle has spent 
there. The method faces though a low energy limit to its accuracy
because the interaction between the particle and the clock perturbs the 
particle and thus the measurement. This limitation 
was formulated by Peres \cite{Peres80,Peres95} as a bound of the time resolution of the clock $\tau$ (the time required to distinguish two orthogonal hand states),
\beq
\label{ine}
\tau E > \hbar,  
\eeq     
for a given particle energy $E$. 
The same result follows by imposing that the transmission
probability be close to 1 \cite{AOPRU};  
it has been also found without reference to any clock from Kijowski's time-of-arrival distribution \cite{Baute}, $\tau$ being in this case the time-of-arrival uncertainty and $E$ the average incident energy; and quite
interestingly, it arises as well as a decoherence condition for consistent histories 
in the analysis of Halliwell and Yearsley of the quantum time of arrival \cite{Halli}.      
If, encouraged by the agreement of so many different approaches, 
we apply this inequality to the energy 
of a Cesium atom moving at speeds of conventional Cs clock standards 
(between $100$ and $300$ m/s), the predicted resolution is $\sim 10^{-14}$ s, 
which is consistent with accuracies achieved in the 70's and 80's 
but it is an order of magnitude too large for more recent conventional Cs clocks \cite{VA05}. If, instead, we assume     
a velocity of 5 cm/s, one of the smallest speeds 
projected so far for a Cs atom clock in space \cite{Laurent}, 
the resulting minimal ``resolution'' is a worryingly high 
0.36 $\mu$s, many orders of magnitude larger than the stated
target accuracy of $10^{-16}$ s for that clock. 
In fact current cold-atom fountain clocks provide accuracies much better than the ones that follow from Eq. (\ref{ine}). 
Clearly the naive application of Eq. (\ref{ine}) is not justified, but 
the question remains as to what the negative effect is, 
if any, of very low atomic motion on atomic clocks.      
 
Quite independently of the works on ``quantum clocks'', time-frequency metrology has experienced a phenomenal progress. Precisely, 
one of the main factors for recent and future 
accuracy improvement is laser cooling and the concomitant use of low atomic velocities \cite{VA05} to make flight times large and the resonance curve 
narrower. Therefore there is indeed a need to evaluate fundamental low energy limits for atomic clocks \cite{SeidelMuga,mou07}.
The characteristic quantum clock process described above (the particle 
activating a handle in a space region),
is similar to the physical basis 
of an atomic clock but there are some important differences
as well: 
in a simple atomic clock
with a one-field (Rabi) configuration, an atom in the ground state 
passes trough a field region in resonance with a
hyperfine atomic transition, which  
induces the internal excitation of the atom. Since the ground or excited state populations
oscillate periodically with Rabi's frequency $\Omega$ when an atom at rest is immersed in the field, it is natural, except possibly for too low energies, 
to relate the passage time to the final internal oscillation phase, which can be read out from the excited population.   
In the atomic clock, 
however, the quantity of interest is not 
really the passage time, even though it plays an important role, 
but the frequency of the transition. This is determined from the central peak of the resonance curve, whose width is inversely proportional to the passage time as we shall see in more detail:  
an analysis 
where the transition of the atom across the field is treated
classically, which is valid when the atomic kinetic energy $E=k^2\hbar/(2m)$ is larger than other relevant energy scales, $E>>\hbar\Delta, \hbar\Omega/2$, ($\Delta$ is the detuning, $m$ the mass,
and $k\hbar/m$ the velocity), shows that the probability of excitation to level $2$, when the incident atom is in the ground state $1$ and the field width is $l$, is given by 
\beq\label{eq:SSE-semiclasica}
P_{12}^{scl}(\Delta)=\frac{\Omega^2}{\Omega^2+\Delta^2}\sin^2\left(\frac{\sqrt{\Omega^2+\Delta^2}T}{2}\right), 
\eeq
%
where  $T=lm/(\hbar k)$ is the (classical) in-field flight-time.  
Three relevant properties of $P_{12}$ are the mentioned  
narrowing of the main peak when $T$ increases, 
the peak at $\Delta=0$ irrespective of
the atomic velocity (or equivalently of $T$),
and the symmetry around $\Delta=0$.
Thus, velocity averaging does not affect the central fringe much, so that  
measurement of $P_{12}$ can be used to steer the frequency of 
an external oscillator close to the reference, 
ideally unperturbed atomic frequency $\omega_0$. This  
steering will be more accurate the 
narrower the peak, i.e., for slower atoms.
The clockwork then produces the second by counting a predetermined
number of oscillations of the external oscillator, $\approx9.2\times 10^9$ for Cs clocks, so that a shift in the peak translates into an error
in the determination of the second of the order of the fractional
frequency error.\footnote{unless of course it is well understood 
and predictable so that it can be corrected by hand.}        

\begin{figure}[t]
\begin{center}
\vspace{1cm}
\includegraphics[height=6.3cm]{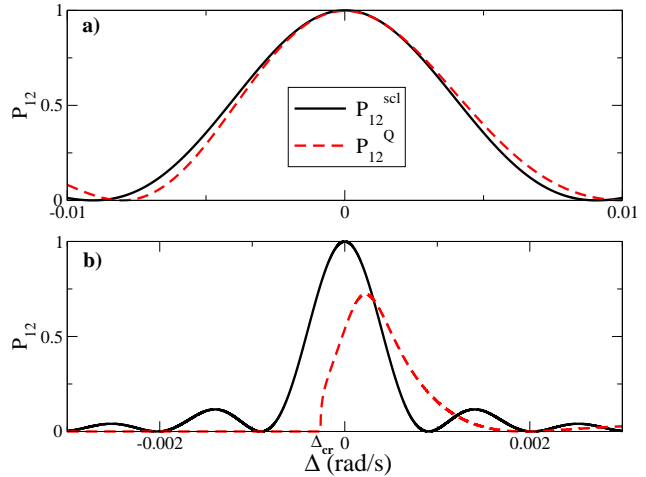}
\caption[]{(Color online) Rabi (one-field) scheme. (a) Perturbative regime: the maximum probability of the exact and semiclassical solutions $P_{12}^Q$ and $P_{12}^{scl}$ practically coincide at $\Delta=0$ for a $\pi$-pulse. Note however the peak asymmetry and the related shift at half height. Data: $l=3$ mm and $v=5$ $\mu$m/s. (b) Reflection-dominated regime: the central fringe is narrower and severely deformed  ($v=0.5$ $\mu$m/s, note the different scale), and starts from a  critical detuning. In (a) and (b) The Rabi frequency is fixed by the $\pi$-pulse condition $\Omega=\pi v/l$. The mass of Cs is used in all figures: $m(Cs)=2.2\times10^{-25}$ Kg.}
\label{resonance}
\end{center}
\end{figure}

Most atomic clocks actually use the Ramsey configuration 
with two separated fields rather than one, to 
improve performance, but the above 
three properties of $P_{12}$ remain true also for the Ramsey scheme.  
In the Ramsey scheme the free-flight time between fields plays the role of $T$
as the parameter to be maximized, within technical constraints, 
for narrowing the resonance.   
We shall first restrict the study to the Rabi configuration for 
simplicity, and provide later the results for the Ramsey configuration, since the manipulations are very similar but lengthier.      
Our main aim is to determine the fractional frequency error of the clock at ``low'' incident velocities: this is the (absolute value of the) shift of the peak of $P_{12}(\Delta)$ with respect to the ideal value 
$\Delta=0$ divided by the atomic transition frequency. 
The term ``low velocity'' here is subtle and must be clarified: we shall consider velocities
where the semiclassical result (with the peak at $\Delta=0$) has to be {\em{corrected}} perturbatively because of quantum motion, but such that the overall 
interference pattern keeps essentially the same form as for the semiclassical limit. 
Physically this means that reflection from the field regions is still very unlikely and $E>>\Omega\hbar$. This will enable us to find the corrections by solving first the quantum dynamics exactly and then, taking the dominant and first correction terms in ``large $k$'' expansions. An example of the perturbative regime is provided in Fig. \ref{resonance}a,
where the quantum curve resembles, but it is slightly perturbed with respect to, the semiclassical curve. We insist that this regime may in fact involve very small 
velocities according to other criteria, such as the velocities in Fig. \ref{resonance}a, 
unrealistically low in any actual atomic clock, but useful for illustrating clearly 
the main features of the quantum curve, in particular the asymmetry, and the maximum very near $\Delta=0$, at least for the naked eye (the maximum is in fact shifted but only slightly as discussed in more detail later on). 
      
Reflection from the field becomes relevant for $E\leq \Omega/\hbar$, but then the interference pattern is drastically deformed. For a Ramsey configuration this implies a multiple scattering scenario with entirely new Fabry-Perot resonances highly sensitive to the velocity
\cite{SeidelMuga} and the existence of a threshold of $P_{12}$ with respect to detuning. This very extreme regime, see an example in 
Fig. \ref{resonance}b, is out the scope of 
current time-frequency metrology so we shall instead concentrate on the perturbative regime.  

%
%

The objective is to evaluate $P_{12}(\Delta)$ quantally for the transmitted atoms. 
We shall use as a simplified model the effective 1D Hamiltonian 
\be\label{eq:hamilt}
\widehat{H} = \frac{\widehat{p}^2}{2m}- \hbar\Delta\ket{2}\bra{2} + \frac{\hbar}{2}\Omega(x)(\ket{1}\bra{2}+\ket{2}\bra{1}),  
\ee
%
which is obtained, see \cite{Lizuain,RM07}, by assuming a semiclassical effective traveling wave field, and applying  
rotating wave and dipolar approximations. The effective microwave field may be realized by two copropagating   
lasers perpendicular to the atomic motion which induce a Raman transition
\cite{raman82,vanier05}. In this case 
the effective detuning $\Delta$ contains, apart from the difference between 
effective laser frequency (difference between the copropagating laser frequencies) and transition frequency, the recoil term $-k_L^2\hbar/(2m)$, $k_L$ being the effective wavenumber (difference between the two laser wavenumbers). 
Finally, $\Omega(x)$ is the effective Rabi frequency that we consider to be 
constant in the illuminated region(s) and zero outside.

%
To find the exact solution of Eq.~(\ref{eq:hamilt}) 
%
we start by solving the stationary Schr\"odinger equation (SSE) for each zone
labeled with an index $\alpha$: 
on the left of the field ($\alpha=I$), in the field ($\alpha=II$), 
and on the right of the field ($\alpha=III$):
\be\label{eq:SSE}
\widehat{H}_\alpha\phi_\alpha=E\phi_\alpha;   (\alpha = I, II, III),
\ee
where  $\phi_\alpha$ is a two-component wavevector 
\be\label{eq:spinordef}
\phi_\alpha=
\left(
\begin{array}{c}
\phi_{\alpha}^{(1)}
\\
\phi_{\alpha}^{(2)}
\end{array}
\right).
\ee
$i)$ For $\alpha=I$,  
%
%
%
\be\label{eq:hamilt-s0}
\left(\frac{\widehat{p}^2}{2m}- \hbar\Delta\ket{2}\bra{2}\right)\phi_{_I}=E\phi_{_I}
\ee
reduces to 
two uncoupled equations for $\phi_{_I}^{(1)}$ and  $\phi_{_I}^{(2)}$.  
The general solution (delta-normalized in $k$ space) is thus  
\beqa\label{eq:SSE-s0-res}
\phi_{_I}&=&\frac{1}{\sqrt{2\pi}}\big[(a_{_I}e^{ikx}+b_{_I}e^{-ikx})\ket{1}
\\\nonumber
&+&(c_{_I} e^{iqx}+d_{_I}e^{-iqx})\ket{2}\big],
\eeqa 
where $q=\sqrt{k^2+2m\Delta/\hbar}$ and  $k=\sqrt{2mE/\hbar^2}$.

$ii)$ For the interaction region, $\alpha=II$, 
%
\be\label{eq:hamilt-s1}
\left[\frac{\widehat{p}^2}{2m}- \hbar\Delta\ket{2}\bra{2}+ \frac{\hbar}{2}\Omega(\ket{1}\bra{2}+\ket{2}\bra{1})\right]\phi_{_{II}}=E\phi_{_{II}}.
\ee
By diagonalizing the interaction part of the Hamiltonian we obtain its wave vectors and eigenstates $\ket{\lambda_{\pm}}$,  
\be\label{eq:autofun-s1}
\ket{\lambda_{\pm}}=
\left(
\begin{array}{clrr}
	\	1	\\
	\frac{2\lambda_{\pm}}{\Omega}\\
\end{array}
\right),
\ee
where $\lambda_{\pm}=\frac{-\Delta\pm\Omega'}{2}$ and $\Omega'=\sqrt{\Delta^2+\Omega^2}$. 

From Eq. (\ref{eq:autovalores-s1}) and  (\ref{eq:autofun-s1}) we may write the solution 
of the SSE in  $\alpha=II$ as 
\beqa\label{eq:SSE-s1-res}
\phi_{_{II}}&=&\frac{1}{\sqrt{2\pi}}\big[(a_{_{II}}e^{ik_+x}+b_{_{II}}e^{-ik_+x})\ket{\lambda_+}
\nonumber\\
&+&(c_{_{II}}e^{ik_-x}+d_{_{II}}e^{-ik_-x})\ket{\lambda_-}\big], 
\eeqa 
with the mode wavenumbers obeying
\be\label{eq:autovalores-s1}
\frac{\hbar^2k_{\pm}^2}{2m}=\frac{\hbar^2k^2}{2m}-\hbar\lambda_{\pm}.
\ee
$iii)$ Finally, for $\alpha=III$ we have, as for $\alpha=I$,  
\beqa\label{eq:SSE-s2-res}
\phi_{_{III}}&=&\frac{1}{\sqrt{2\pi}}\big[(a_{_{III}}e^{ikx}
+b_{_{III}}e^{-ikx})\ket{1}
\nonumber\\
&+&(c_{_{III}}e^{iqx}+d_{_{III}}e^{-iqx})\ket{2}\big].
\eeqa 
In all cases, $\alpha=I,II,III$, it is useful to introduce the  
vector of coefficients   
\beq
\boldsymbol{v}_{\alpha}=(a_{\alpha},b_{\alpha},c_{\alpha},d_{\alpha})^{\cal T}, 
\eeq
where ${\cal T}$ means ``transpose''. 
\section{Transfer matrices and excitation probability}
Let us consider a plane wave $\frac{1}{\sqrt{2\pi}}e^{ikx}$, incident from the left  ($x\rightarrow-\infty$), with the atoms in the ground state  $\ket{1}$ so that  $a_{_I}=1$, $c_{_I}=0$, $b_{_{III}} =d_{_{III}}=0$. After the interaction ``barrier'', the transmitted atom traveling to $x\rightarrow\infty$ may stay 
in $\ket{1}$ or be excited in $\ket{2}$, so that
\beqa\label{eq:v0}
\boldsymbol{v}_{_I}&=&(1,r_{11},0,r_{12})^{\cal T},
\nonumber\\
%
%
\label{eq:v2}
\boldsymbol{v}_{_{III}}&=&(t_{11},0,t_{12},0)^{\cal T},
\eeqa
where $r_{ij},t_{ij}$ are reflection and transmission amplitudes for left incidence 
in internal state $i=1,2$ and outgoing internal state $j=1,2$.     

%
%
Let us now introduce for the outer regions, $\alpha=I,III$,
a vector containing the amplitudes and their derivatives,   
\begin{eqnarray}
\left(
\begin{array}{c}
\phi_\alpha^{(1)}(x)\\
\phi_\alpha^{(2)}(x)\\
\dot{\phi}_\alpha^{(1)}(x)\\
\dot{\phi}_\alpha^{(2)}(x)
\end{array}
\right)=M_0(x)\boldsymbol{v}_{\alpha},
\end{eqnarray}
and similarly, inside the field ($\alpha=II$)
\begin{eqnarray}
\left(
\begin{array}{c}
\phi_{_{II}}^{(1)}(x)\\
\phi_{_{II}}^{(2)}(x)\\
\dot{\phi}_{_{II}}^{(1)}(x)\\
\dot{\phi}_{_{II}}^{(2)}(x)
\end{array}\right)=M_b(x,\phi_1)\boldsymbol{v}_{_{II}},
\end{eqnarray}
where the dot represents derivative with respect to $x$. 
The explicit form of the matrices $M_0(x)$ and  $M_b(x)$ is  
\be\label{eq:M0}
M_0(x)=\frac{1}{\sqrt{2\pi}}
\left(
\begin{array}{clrr}
	e^{ikx}&e^{-ikx}&0&0\\
	0&0&e^{iqx}&e^{-iqx}\\
	ike^{ikx}&-ike^{-ikx}&0&0\\
	0&0&iqe^{iqx}&-iqe^{-iqx}\\
\end{array}
\right)
\ee
\begin{widetext}
\be\label{eq:m}
M_b(x)=\frac{1}{\sqrt{2\pi}}
\left(
\begin{array}{clrr}
e^{ik_+x}  &   e^{-ik_+x}  &   e^{ik_-x} &e^{-ik_-x}
\\
\frac{2\lambda_+e^{ik_+x}}{\Omega}&\frac{2\lambda_+e^{-ik_+x}}{\Omega}&\frac{2\lambda_-e^{ik_-x}}{\Omega}&\frac{2\lambda_-e^{-ik_-x}}{\Omega}
\\
ik_+e^{ik_+x}&-ik_+e^{-ik_+x}&ik_-e^{ik_-x}&-ik_-e^{-ik_-x}
\\
\frac{2ik_+\lambda_+e^{ik_+x}}{\Omega}&\frac{-2ik_+\lambda_+e^{-ik_+x}}{\Omega}&\frac{2ik_-\lambda_-e^{ik_-x}}{\Omega}&\frac{-2ik_-\lambda_-e^{-ik_-x}}{\Omega}
\end{array}
\right)
\ee
\end{widetext}
The matching conditions at $x_1$ and  $x_2$ for the wave functions 
and their derivatives can now be written as 
%
%
%
%
\be\label{eq:una}
M_0(x_1)\boldsymbol{v}_{_I}=M_b(x_1)\boldsymbol{v}_{_{II}},
\ee
\be\label{eq:dos}
M_b(x_2)\boldsymbol{v}_{_{II}}=M_0(x_2)\boldsymbol{v}_{_{III}}.
\ee
Eliminating $\boldsymbol{v}_{_{II}}$ from the system above,
we end up with a transfer matrix $\textsf{T}(x_1,x_2,\phi_1)$ which connects
the amplitude vectors of both sides, 
\begin{equation}
\boldsymbol{v}_{_I}=\textsf{T}(x_1,x_2)\boldsymbol{v}_{_{III}},
\end{equation}
and is given by the matrix product
\begin{equation}
\textsf{T}(x_1,x_2)=M_0(x_1)^{-1}M_b(x_1)M_b(x_2)^{-1}M_0(x_2).
\label{tram}
\end{equation}
Using Eqs.~(\ref{eq:v2},\ref{eq:M0},\ref{eq:m}) 
we may calculate the transmission amplitude for passing from $\ket{1}$ to $\ket{2}$
from the left as 
\be\label{eq:t12}
t_{12}=\frac{\sa{T}_{31}}{\sa{T}_{31}\sa{T}_{13}-\sa{T}_{33}\sa{T}_{11}}.
\ee
The exact quantum excitation probability is thus, generalizing the semiclassical result (\ref{eq:SSE-semiclasica}),    
\be\label{eq:prob}
P_{12}^{Q}(\Delta)=\frac{q}{k}\left|t_{12}\right|^2.
\ee
%

\section{Quantum Motion Shifts}
Two different shifts will be considered, $\Delta_Q^{(M)}$ and $\Delta_Q^{(hh)}$:
the first one is the shift of the maximum 
of the resonance peak with respect to $\Delta=0$; the second one is the shift of the 
average between the resonance curve points at half height. As hinted by Fig. \ref{resonance}a they
have quite a different behavior,  
the first one being negligible for the naked eye whereas the second one is clearly significant. 
We shall provide analytical expressions 
for the former, and numerical results for the later.

%

We shall first calculate the maximum value of $P_{12}^{Q}(\Delta)$. From  
\be\label{eq:maximo}
\frac{dP_{12}^{Q}}{d\Delta}=0
\ee   
and Eq. (\ref{eq:prob}), the maximum must satisfy 
%
%
%
\be\label{eq:maximo1}
t_{12} {t_{12}}^*+\frac{q^2\hbar}{m}\left(\frac{dt_{12}}{d\Delta}{t_{12}}^*+t_{12} \frac{d{t_{12}}^*}{d\Delta}\right)=0.
\ee
The explicit expression in terms of the basic variables $v,L,\Omega$
is long and cumbersome,
since each element ${\textsf T}_{ij}$ of the matrix ${\textsf T}(x_1,x_2)$ in Eq. (\ref{eq:t12}) 
results from multiplying four matrices $4\times4$.
To simplify we shall first eliminate $\Omega$ by imposing 
$\Omega=\frac{\hbar k \pi}{lm}$, which is the semiclassical condition 
for a $\pi$-pulse at $\Delta=0$,  
%
and then expand $t_{12}$ in powers of $\Delta$  
%
%
up to second order,  
%
%
%
%
%
%
\be\label{eq:t12aprox1-de}
t_{12,\Delta}\approx\gamma_0(k)+\gamma_1(k)\Delta+\gamma_2(k)\Delta^2, 
\ee
where  
\beqa\label{eq:gamma0}
\gamma_0&=&\left[t_{12}\right]_{\Delta=0}
\\
\label{eq:gamma1}
\gamma_1&=&\left[\frac{\partial{t_{12}}}{\partial{\Delta}}\right]_{\Delta=0}
\\
\label{eq:gamma2}
\gamma_2&=&\frac{1}{2}\left[\frac{\partial^2{t_{12}}}{\partial{\Delta^2}}\right]_{\Delta=0}.
\eeqa
The probability $P_{12}^{Q}(\Delta)$ is approximately  
\be\label{eq:p12aprox-de}
P_{12}^{Q}(\Delta)=\frac{q}{k}\left|
(\gamma_0+\gamma_1 \Delta + \gamma_2\Delta^2)\right|^2.
\ee
%

\begin{figure}[t]
\begin{center}
\vspace{1cm}
\includegraphics[height=6.3cm]{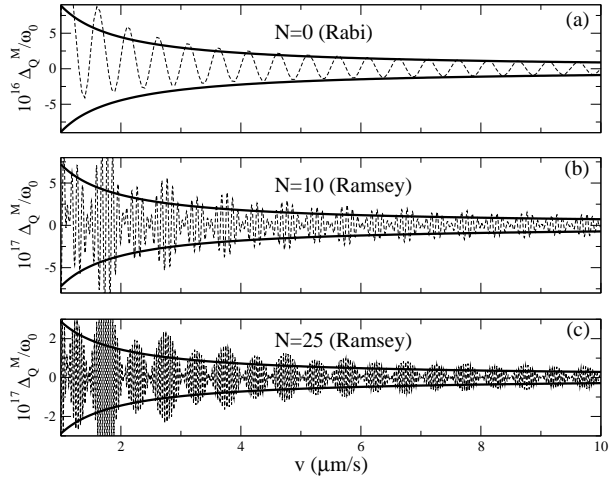}
\caption[]{Comparison between numerical (dashed lines) and analytical values (solid lines), of $\Delta_Q$ over the transition frequency $\omega_0$ for a Cs atom at different velocities. 
(a) Single Rabi $\pi$-pulse with $l=3$ mm, Rabi frequency fixed by $\pi$-pulse condition, $\Omega=\pi v/l$. For the analytical results only upper/lower bounds of Eq. (\ref{eq:laformula}) have been plotted. The approximations are better for higher velocities.
(b) and (c) Ramsey $\pi/2$ pulses with $l=1.5$ mm and separated by $L$, ($N=L/l$), Rabi frequency fixed by $\pi/2$-pulse condition, $\Omega=\pi v/2l$.}
\label{fig_02}
\end{center}
\end{figure}

Substituting the expansions in Eq.~(\ref{eq:maximo1}),  
\begin{multline}
\left|(\gamma_{0}+\gamma_{1}\Delta+\gamma_2\Delta^2)\right|^2+
\\
2q^2\frac{\hbar}{m}$Re$\left[(\gamma_1+2\gamma_2\Delta)(\gamma_{0}+\gamma_{1}\Delta
+\gamma_2\Delta^2)^*\right]=0.
\end{multline}
Gathering terms, truncating in first order of $\Delta$ and using 
\begin{eqnarray}
\nonumber
\theta_0(k)&=&\gamma_0\gamma_0^*+q^2\frac{\hbar}{m}(\gamma_1\gamma_0^*+\gamma_0\gamma_1^*)
\\\nonumber
\theta_1(k)&=&(\gamma_1\gamma_0^*+\gamma_0\gamma_1^*)
+2q^2\frac{\hbar}{m}(\gamma_1\gamma_1^*+\gamma_2\gamma_0^*+\gamma_0\gamma_2^*)
\nonumber
\end{eqnarray}
the peak must satisfy   
\begin{eqnarray}
\nonumber
\theta_0(k)+\theta_1(k)\Delta+{\cal O}(\Delta^2)=0,
\nonumber
\end{eqnarray}
which can be solved retaining the dominant order in $k$ 
to find the explicit shift expression     
\be\label{eq:laformula}
\Delta_{Q}^{(M)}(k)=\frac{\hbar^2\pi^4}{16m^2vl^3}
\sin\left(2 k l\right),
\ee
where $v=k\hbar/m$ is the velocity, 
see Fig.~\ref{fig_02}a, and the superscript 
$M$ stands for ``maximum''.
%

%
%
 
We may proceed similarly for the Ramsey configuration and  
write down a new transfer matrix which is of course more complicated
but in principle explicit with the aid of a program of 
algebraic manipulation. This new transfer matrix depends on the previous parameters and on $L$,  
the length between fields.  
For ``large'' $k$, a large ratio $N=L/l$, and imposing $\pi/2$ pulses 
at each field, now $\Omega=\frac{\hbar k \pi}{2lm}$, we find the approximate 
envelope 
\beq
\label{DelRamsey}
(\Delta_Q^{(M)})_{env}=\pm\frac{\hbar^2\pi^2}{16 m^2 v l^3 N}.
\eeq
Note that the Ramsey shifts are considerably smaller
than Rabi shifts because of the $N$ in the denominator.
Compared to the Rabi case, the shift for the Ramsey configuration 
oscillates with respect to $k$ with a short period $\pi/L$,
superimposed to the long period $\pi/l$,  
see Fig. \ref{fig_02}b,c.    
These shifts are in any case rather small and, in addition, they oscillate rapidly with respect to $v$ so that they will
average out because of the velocity spread of the atomic cloud.  
%

There is however a second type of shift which is quite relevant in practice: it is defined as the average of the resonance curve detunings at half height of the central peak, $\Delta_Q^{(hh)}$.
We numerically find out that 
the resonance peak is asymmetric and that, at variance with the former  
definition, this shift does not oscillate around zero but it tends to a constant at ``large'' $k$ for fixed $l$ and $L$. 

\begin{figure}
\epsfxsize=8cm
\includegraphics[width=\linewidth]{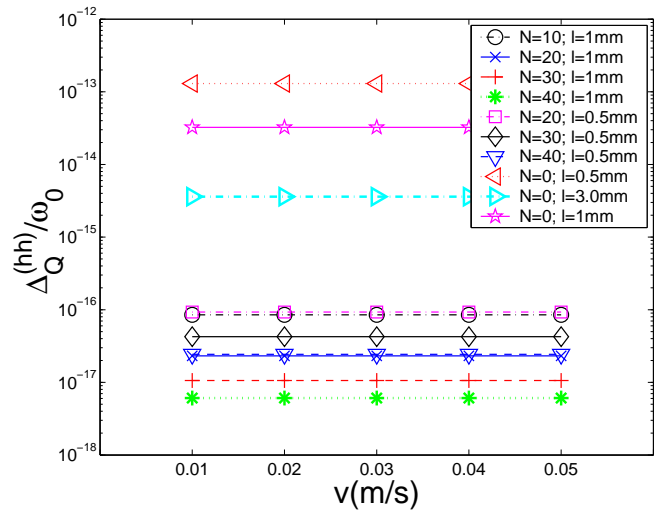}
\caption{(Color online) Constant values of the fractional frequency offset
for the Rabi ($N=0$) and Ramsey
configurations versus incident velocity for different values 
of $l$ and $N=L/l$. Exact results.}
\label{Ramsey1a5}
\end{figure}

\begin{figure}
\epsfxsize=8cm
\includegraphics[width=\linewidth]{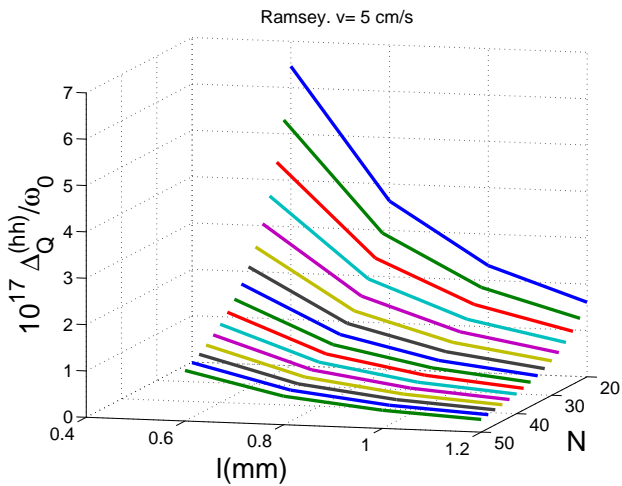}
\caption{(Color online) Fractional frequency offset 
versus $l$ and $N=L/l$ for $v=5$ cm/s. Exact results}
\label{3d}
\end{figure}

\begin{figure}
\epsfxsize=8cm
\includegraphics[width=\linewidth]{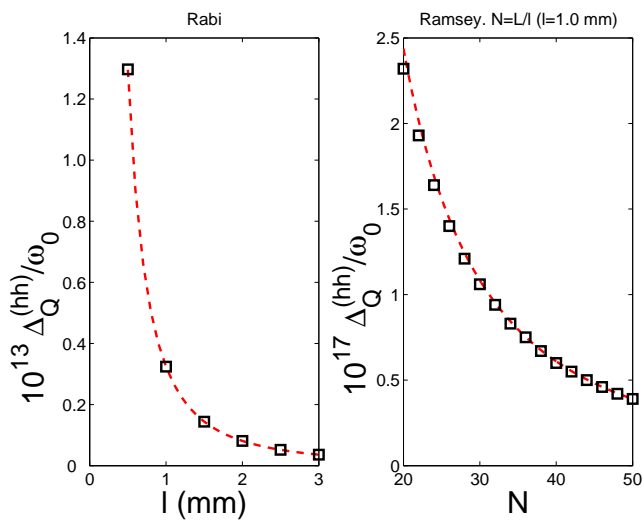}
\caption{(Color online) Fractional frequency offset at half height  
versus $l$ and $N=L/l$ for $v=5$ cm/s. Exact results(squares) and fits to an inverse square form (dashed).}
\label{hhrabi}
\end{figure}

As shown e.g. in Fig. \ref{resonance}a, this shift is more significant than the former, both because of its larger magnitude and because of the 
absence of oscillation with respect to velocity, see Fig. \ref{Ramsey1a5}. 
Note that the quantum excitation curve tends to the semiclassical curve when compared ``vertically'' for increasing $v$ and for a given $\Delta$, $P_{12}^Q-P_{12}^{scl}\to 0$. This is compatible with the fact that, when 
compared ``horizontally'' at half height, the two curves are separated by a shift which stays constant when increasing $v$. If we represent only the central peak, 
this constant shift is less and less visible 
due to the peak broadening and changing scale when $v$ increases.     
A back-of-the-envelop argument may help to gain some intuitive understanding of this 
phenomenon giving relevant dependences.
First notice that the wavenumber $q$ is not an even function with respect to $\Delta$
and provides asymmetry: the excited atom finds a lower potential ground with positive detuning than with negative detuning and the quantum excitation probability at 
half height may be approximated as $(q/k)P^{scl}$, see Eq. (\ref{eq:prob}), or simply $q/(2k)$, larger than 1/2 for positive detuning and smaller otherwise. Expanding for large $k$, and approximating $\Delta$ as half the distance to the first zero, the vertical difference between quantum and semiclassical curves at 
half height is, for the Rabi scheme, $\approx\pi/(4 k l)$. On the other hand the slope, again estimated from the zero and for the Rabi case is $\approx lm/(\hbar k \pi)$. As this slope is also the ratio between vertical and horizontal shifts, 
there results a $k$-independent horizontal shift proportional to $l^{-2}$ for Rabi excitation\footnote{The full predicted shift is $\hbar\pi^2/(4ml^2)$ which provides remarkably good results given the roughness of the argument.}, or to $L^{-2}$ for Ramsey excitation and large $N$. This dependences are confirmed in Figs. \ref{Ramsey1a5}, 
\ref{3d}, and \ref{hhrabi}. 
%

%
%
%
\section{Conclusions}
The bound for the time resolution of a quantum clock, Eq. (1),  suggests 
limitations to the accuracy of atomic clocks at low atomic 
velocities. 
One of the derivations of Eq. (1) relies on the reflection of particles from the interacting field.  
For an atomic beam crossing a field near resonance this means   
$E>>\Omega\hbar$, which is reminiscent of Peres' bound. However $1/\Omega$ does not play at all the role of the time resolution of an  atomic clock, which is
nowadays even much better than the oscillation period of the atomic transition, 
a modest $10^{-10}$ Hz. Motivated by the facts that a naive application of Eq. (1) does not provide a valid assessment of actual low energy limits
of atomic clocks,   
and that nevertheless, the trend to use lower and lower atomic velocities requires this type of information, we have investigated 
resonance shifts in atomic clock configurations of Rabi and Ramsey type
due to quantum atomic motion.  
Explicit expressions are provided for the shift of the maximum, 
Eqs. (\ref{eq:laformula}) and (\ref{DelRamsey}), but this effect is quite small compared to the  
peak asymmetry, and moreover it would cancell out because of averaging over incident velocities. Instead, when the resonance is defined by the middle point of the resonance peak at half height, the peak asymmetry implies a   
significant shift, several orders of magnitude larger than the shift of the maximum,
and robust with respect to momentum averaging.    
Both shifts are smaller in the Ramsey scheme than in the Rabi scheme, but even so the half-height asymmetry shift 
would have to be considered in the error budget of  
accurate Raman-Ramsey clocks. It can be reduced by increasing the 
width of the interaction region between field and atom, and, in the Ramsey scheme, by increasing the free-flight length. The present results motivate further work to investigate similar effects in standing wave rather than traveling wave configurations
and for smooth field intensities. 

Finally, we would like to comment on  
atomic clock designs based on trapped ions and
illumination pulses in time rather than fixed in space.
As the ion is trapped, it might seem that such a scheme is
free from quantum-motion shifts but this is not the case. 
We have provided expressions for the corresponding
shifts in \cite{LME}.


%
%


\begin{acknowledgments}
We are very grateful to R. Wynands for 
useful comments on the manuscript.  
This work has been supported by Ministerio de Educaci\'on y Ciencia (FIS2006-10268-C03-01) and the Basque Country University (UPV-EHU, GIU07/40). 
\end{acknowledgments}
%
%

\end{document}